\documentstyle[prb,psfig,aps]{revtex}

\newcommand{\be}{\begin{equation}} \newcommand{\ee}{\end{equation}}
\newcommand{\bt}{\begin{tabular}} \newcommand{\et}{\end{tabular}}

\begin{document}

\draft

\twocolumn[\hsize\textwidth\columnwidth\hsize\csname@twocolumnfalse%
\endcsname

\title{
The Two-Dimensional Disordered Boson Hubbard Model: Evidence for a
Direct\\
Mott Insulator-to-Superfluid Transition and 
Localization in the Bose Glass Phase}

\author{Jens Kisker} \address{Institut f\"ur Theoretische Physik,
  Universit\"at zu K\"oln, D-50937 K\"oln, Germany}

\author{Heiko Rieger} \address{HLRZ c/o Forschungszentrum J\"ulich,
  D-52425 J\"ulich, Germany}

\date{\today}

\maketitle

\begin{abstract}
  We investigate the Bose glass phase and the insulator-to-superfluid
  transition in the two-dimensional disordered boson Hubbard model in
  the Villain representation via Monte Carlo simulations. In the Bose
  glass phase the probability distribution of the local susceptibility
  is found to have a $1/ \chi^2$ tail and the imaginary time Green's
  function decays algebraically $C(\tau) \sim \tau^{-1}$, giving rise
  to a divergent global susceptibility. By considering the
  participation ratio it is shown that the excitations in the Bose
  glass phase are fully localized and a scaling law is established.
  For commensurate boson densities we find a {\em direct} Mott
  insulator to superfluid transition without an intervening Bose glass
  phase for weak disorder.  For this transition we obtain the critical
  exponents $z=1, \nu=0.7\pm 0.1$ and $\eta = 0.1 \pm 0.1$, which
  agree with those for the classical three-dimensional XY model
  without disorder. This indicates that disorder is irrelevant at the
  tip of the Mott-lobes and that here the inequality $\nu\ge2/d$ is
  violated.
\end{abstract}


] \newpage

\section{Introduction}
The localization transition in disordered systems of interacting bosons 
has attracted a lot of interest in recent years. The observation is that 
there is a direct superconductor to insulator transition at zero temperature,
which can be tuned by varying a suitable control parameter like
the disorder strength or an applied magnetic field. 
The paradigmatic two-dimensional realizations of such systems 
are amorphous superconducting films \cite{liupalaanen}, where
the Cooper pairs are considered as bosons, and ${\rm ^4He}$ adsorbed in porous 
media \cite{crowell}. On the theoretical side, these systems are described
most simply by a boson Hubbard model, and a large amount of effort has been 
put forth to study the various aspects of this transition.
This includes mean-field calculations \cite{fwgf}, 
real space renormalization group studies \cite{singh}, strong coupling 
expansions \cite{freericks} and
quantum Monte Carlo investigations in one- \cite{scalettar} and two
dimensions \cite{soerensen,krauth,otterlo}. 

Of particular interest are the properties of the so called Bose glass
phase which is supposed to emerge for strong disorder, and which is 
characterized by being insulating but compressible and gapless \cite{fwgf}. 
The latter property leads to a diverging superfluid susceptibility throughout 
the Bose glass phase, which is reminiscent of the quantum Griffiths phase in 
random transverse Ising systems \cite{griffiths,dsfisher,young,rieger,guo}. 
It is one of the main goals of this work to investigate the Bose glass phase 
and to characterize the Griffiths singularities by means of Monte Carlo 
simulations of the 2D Bose Hubbard model. The other major point of interest 
is the insulator to superfluid transition for weak disorder
and commensurate boson densities, since there is an ongoing debate whether 
this transition occurs from the Mott insulating (MI)- or the Bose glass 
(BG) phase \cite{fwgf,singh,krauth}. 
In this paper we present numerical evidence for a {\em direct} Mott
insulator to superfluid transition that falls in the universality class
of the pure classical XY-model in three dimensions. A short account of
our results has been given recently \cite{kisker}.

The paper is organized as follows: In Sec. \ref{secmodel} the 2D Bose
Hubbard model is introduced and it is briefly stated, how one arrives
via a set of transformations at a 3D classical model, which is used
subsequently in the simulations. In Sec. 
\ref{secscalingth} we introduce the quantities of interest and summarize
the scaling theory of Ref. \cite{fwgf} which is needed for a finite size
scaling analysis of the data. In Sec. \ref{secmc} the Monte Carlo
algorithm is presented. In Sec. \ref{secbg} we examine the Bose glass phase 
for strong disorder and in Sec. \ref{secweakdis} we consider the case of 
weak disorder and address the question of a direct MI-SF transition.
\section{The Model}
\label{secmodel}
We consider the boson Hubbard model with disorder on a square 
lattice in $d=2$ dimensions. The Hamiltonian is given by
\be
H_{BH} = H_0 + H_1
\label{bh}
\ee
\be
H_0 = -t \sum_{\langle i,j \rangle} ({\hat \Phi}_i^{\dagger} 
         {\hat \Phi}_j +{\hat \Phi}_i {\hat \Phi}_j^{\dagger} )
\ee
\be
H_1 = \frac{U}{2} \sum_i {\hat n}_i^2 - \sum_i (\mu + v_i){\hat n}_i \;,
\ee
where ${\hat \Phi}_i$ $({\hat \Phi}_i^{\dagger}$) is the boson annihilation 
(creation) operator at site $i$ and 
${\hat n}_i={\hat \Phi}_i^{\dagger}{\hat \Phi}_i$ 
is the usual number operator. The hopping matrix is $t$ for
nearest neighbors and zero otherwise. The bosons interact via an
on-site repulsion $U$, $\mu$ is the chemical potential and
$v_i$ represents the random
on-site potential varying uniformly in space between $-\Delta$ and $\Delta$.

As a simplification we take the ''phase-only'' approximation 
\cite{qphmfisher}, where amplitude fluctuations in (\ref{bh})
are integrated out. One assumes that the complex field $\Phi$ has the simple
form $\Phi_i = |\Phi_0| e^{i{\hat \phi}_i}$, where ${\hat \phi}_i$ is the phase
operator conjugate to the number operator ${\hat n}_i$ 
 with commutation relation $[\phi_i,n_j]=i\delta_{ij}$.
This yields the quantum-phase-Hamiltonian
\be
H_{QPH} = H_0 + H_1
\label{qph}
\ee
\be
H_0 = -t \sum_{\langle i,j \rangle} \cos({\hat \phi}_i - {\hat \phi}_j) 
\ee
\be
H_1 = \frac{U}{2} \sum_i {\hat n}_i^2 - \sum_i (\mu + v_i) {\hat n}_i \;,
\ee
where ${\hat n}_i$ now measures the deviation from a mean density.

\subsection{Duality transformation}
Through a sequence of transformations the 2D quantum mechanical
Hamiltonian (\ref{qph}) can be mapped onto a classical model of
divergence-free integer link variables in $(2+1)$ dimensions 
which is suitable for Monte Carlo simulations \cite{soerensen}. 

The transformation uses the standard Suzuki-Trotter \cite{suzuki} formula 
where the inverse temperature $\beta$ is divided into $L_\tau$ time slices of 
width $\Delta \tau = \beta / L_\tau$ yielding a path integral representation
of Hamiltonian (\ref{qph}). Taking the Villain approximation 
\cite{villain}, which replaces the exponentiated cosine term in the 
partition function by a sum of periodic gaussians, and after a further 
Poisson summation, one can show 
that the ground state energy density of the quantum model (\ref{qph}) is
equal to the free energy density of a purely classical model
\be
f = \lim_{T\rightarrow 0} \frac{T}{L^2} \ln \mbox{Tr} \; e^{-\beta H_{QPH}} = 
                       - \frac{1}{L^2 L_{\tau} \Delta \tau} \ln 
                        \mbox{Tr} \; e^{-H_{\rm class}/K} \; ,
\label{freeenergy}
\ee
where the classical Hamiltonian $H_{\rm class}$ is given by 
\be
H_{\rm class} = \sum^{\nabla \cdot {\bf J} = 0}_{(i,\tau')} 
 \left(   \frac{1}{2} {\bf J}^2_{i,\tau'} - (\mu + v_i)J^{\tau}_{i,\tau'}
   \right) \; .
\label{action}
\ee 
and the parameter $K$ acts as an effective temperature of the model
and is used to tune through the phase transition. $K$ corresponds
to $t/U$ in the quantum phase model (\ref{qph}) and thus by changing
$K$ one moves horizontally, i.e. parallel to the $t/U$ axis in the
phase diagram of Hamiltonian (\ref{qph}). The ${\bf J}'s
=(J^x,J^y,J^{\tau})$ are three-component vectors consisting of integer
variables on the links of the original lattice running from $-\infty$
to $\infty$. The $\tau$-component $J^\tau$ corresponds to the particle
number. Note that the summation in (\ref{action}) is just over {\em
  divergence-free} configurations of the link variables.  Strictly
speaking, one has to take the limit $L_\tau \rightarrow \infty$ for
the equality in eq. (\ref{freeenergy}) to be valid. In the simulations
however, we have to keep $L_\tau$ finite, which means that we are
working at a temperature $T \sim 1/ L_\tau$ for the original quantum
phase model (\ref{qph}).

The action (\ref{action}) describes an ensemble of string like objects in a 
random potential, and considering the divergence-free condition, these 
objects can be interpreted as flux-lines in a high-$T_C$ superconductor.
This observation is not surprising, since one has the general correspondence
between 2D bosons at zero temperature and flux-lines in 3D 
high-$T_C$ superconductors \cite{fisher2d,ceperley,feynman}.

\subsection{Phase diagram}
\label{phasediagram}

The phase diagram of the classical Hamiltonian (\ref{action}) in the
thermodynamic limit ($L\to\infty$, $L_\tau\to\infty$), which
corresponds to the $T=0$ phase diagram of the quantum model
(\ref{qph}) can be obtained qualitatively by simple perturbative
arguments \cite{fwgf}.

For the case without disorder ($v_i=0$) one obtains two phases: a Mott
insulating and a superconducting phase. In the $\mu$ vs. $K$ plane the
Mott insulating phase has a lobe like shape which can be understood by
first considering the atomic limit $K=0$ and then allow for small
hopping in the quantum phase model (\ref{qph}), i.e.\ finite effective
temperature for the classical model (\ref{action}). At zero effective
temperature, every site is occupied by an integer number $n=J^\tau$ of
bosons which minimizes the on site energy $\epsilon(n) = - \mu n +
\frac{1}{2}n^2$. Thus in the interval $n < \mu + \frac{1}{2} < n+1$
the occupation number is fixed to $n$ at every site causing the
compressibility $\kappa = \partial n / \partial \mu$ to vanish.
Furthermore there is a gap in the single particle spectrum making the
Mott phase indeed an insulator. 

For increasing hopping amplitude $t/U$ the bosons in the original
quantum phase model (\ref{qph}) can gain kinetic energy by hopping on
the lattice and thus the gap decreases with increasing $t$. When the
gap vanishes, the bosons are free to hop on the lattice thus producing
superfluidity. Equivalently in the classical model (\ref{action}) an
increase in effective temperature $K$ causes a distortion of the
initially straight world lines allowing them to move around freely
above some critical value for $K$ giving the Mott phase a lobe like
shape.  As Hamiltonian (\ref{action}) is periodic in $\mu$ with period
1, the lobes are repeated on the $\mu$ axis.

In the presence of disorder a new, insulating Bose glass phase is
supposed to emerge \cite{fwgf} in addition to the Mott insulating and
the superfluid phase. As the disorder strength $\Delta$ increases, the
Mott lobes shrink since there will always be sites where the energy
necessary to add a particle or hole is reduced due to the disorder and
these sites therefore favour the addition of particles or holes. Thus
the compressibility ceases to vanish and the trace of the Mott
insulator disappears. Fisher {\em et al.} now argue that the extra
bosons will not immediately produce superfluidity but will still be
localized due to the disorder. Thus the new phase is characterized by
a non-vanishing compressibility, no energy gap and zero conductivity.
Only for larger hopping amplitudes the bosons will become delocalized
resulting in a superfluid phase.

The phase diagram for weak disorder suggested by our simulations is depicted 
in Fig. \ref{figphaseweak}. Note that contrary to Ref. \cite{fwgf} 
we find a {\em direct} MI-SF transition for integer values of the chemical 
potential (see Sec. \ref{secweakdis}) and therefore there is no intervening
BG phase at the tip of the lobe.
\begin{figure}
\psfig{file=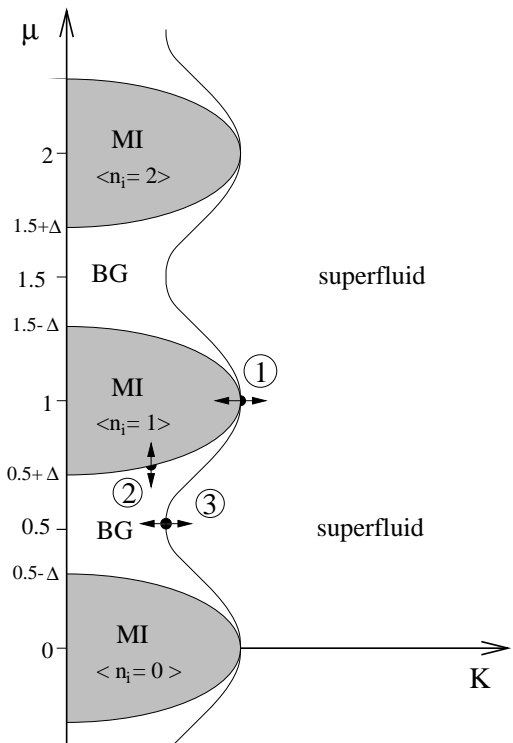,width=\columnwidth}
\caption{The $L_\tau=\infty$ phase diagram of the classical model
  (\ref{action}) for weak disorder suggested by our results presented
  in this paper.  The Mott lobes are centered around integer values of
  the chemical potential.  For integer values of $\mu$ we find a {\em
  direct} MI-SF transition at a multicritical point, therefore there
  is no intervening BG phase at the tip of lobe. Since the Hamiltonian
  (\ref{action}) is periodic in the chemical potential with period one,
  the Mott lobes are repeated in the y-direction. The dots denote
  the phase transitions we examine in the simulations and are labeled
  for later reference. The arrows indicate whether $\mu$ or $K$ 
  were used to tune through the different transitions.}
\label{figphaseweak}
\end{figure}

For disorder strengths $\Delta \ge U/2$ one obtains the phase diagram shown
in Fig. \ref{figphasestrong}. The Mott lobes have totally disappeared
and only the BG- and the superfluid phase remain. 
\begin{figure}
\psfig{file=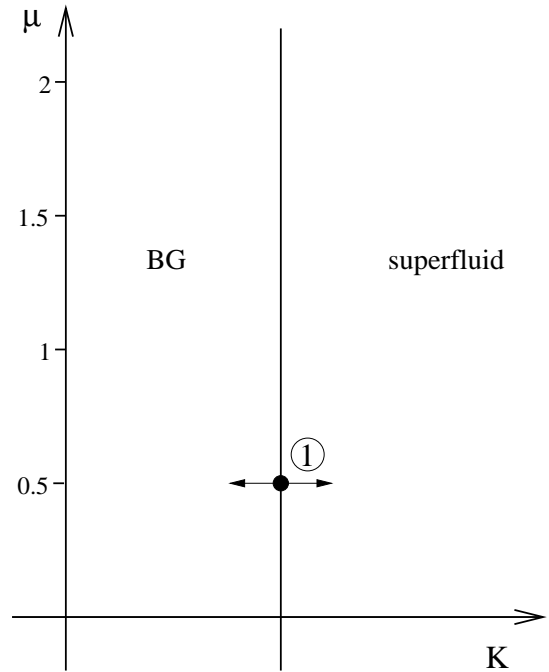,width=\columnwidth}
\caption{The zero temperature phase diagram of the classical model
(\ref{action}) for disorder strengths $\Delta \ge 1/2$. The Mott
lobes have disappeared and only the BG- and the SF-phase remain. The dot
denotes the BG-SF transition studied in Sec. \ref{secbg} and is
identical to point \#3 in Fig. \ref{figphaseweak}.}
\label{figphasestrong}
\end{figure}

Furthermore Fisher {\em et al.} argue that due to the non-vanishing density of
states at zero energy the imaginary time Green's function $C(\tau)$ decays 
algebraically in the BG phase. This leads to a divergent superfluid
susceptibility $\chi = \int C(\tau) \, d\tau$ in the BG phase, which thus
resembles the quantum Griffiths phase found in other disordered systems 
\cite{griffiths,dsfisher,young,rieger,guo}. We investigate this scenario in
Sec. \ref{secbg}.

\section{Scaling Theory and finite-size scaling}
\label{secscalingth}
For completeness we give a brief summary of the $T=0$ scaling theory 
developed in Ref. \cite{fwgf}.
Furthermore we introduce the quantities of interest and show how they can 
be expressed in terms of the link variables of the classical model 
(\ref{action}). Since in the simulations we are working with finite systems, 
it is also shown how to obtain results via finite-size scaling.
\subsection{Stiffness}
\label{scalingstiffness}
From the anisotropy of the classical model (\ref{action}) it can be seen, 
that near a continuous phase transition there are two diverging correlation
"lengths" $\xi$ and $\xi_{\tau}$. This is due to the fact that we are 
considering a $T=0$ quantum phase transition where static and 
dynamic properties of a system are linked. $\xi$ is the correlation
length in the spatial directions, whereas $\xi_{\tau}$ denotes a
diverging time scale in the imaginary time direction. As usual we
introduce the critical exponent $\nu$ by $\xi \sim \delta^{-\nu}$.
Generally $\xi_{\tau}$ will diverge with a different exponent so 
we define the dynamical critical exponent $z$ by $\xi_{\tau} \sim
\xi^z$. The control parameter $\delta$ measures the normalized distance 
from the critical point, which is
$\delta=(K-K_C)/K_C$ or $\delta = (\mu - \mu_C)/\mu_C$ in the simulations.

An important quantity in the further analysis is the zero frequency
stiffness $\rho(0)$,
which is proportional to the order parameter, i.e. the condensate. Thus 
by calculating the stiffness we are able to locate the critical point of 
a phase transition from an insulating to a superconducting phase.
The stiffness measures the change of the free energy under imposing
a twist $\theta$ in the phase of the order parameter and is defined by
\be
\frac{\delta f_s}{\hbar } = \frac{1}{2} \rho(\nabla \theta)^2  \; ,
\ee
where $f_s$ denotes the singular part of the free energy. Via hyperscaling
the scaling form 
\be
\rho \sim \xi^{-(d+z-2)}
\label{scalexi}
\ee
can be derived \cite{fwgf}.

To calculate the stiffness in the simulations we have to express $\rho$ in 
terms of the link variables of the classical model (\ref{action}). By 
considering the effect of an external vector potential one
obtains \cite{soerensen}
\be
\rho(0) = \frac{1}{L_{\tau}} [\langle n_x^2 \rangle ]_{\rm av} \; ,
\label{linkstiffness}
\ee
where $n_x$ defines the winding number
\be
n_x = \frac{1}{L} \sum_{(i,\tau)} J^x_{(i,\tau)} \; ,
\ee
and $[\ldots]_{\rm av}$ denotes a disorder average.
Since at the critical point the diverging correlation lengths are much larger
than the system size we need to obtain
a finite-size scaling form of the stiffness to analyse our data. Because
there are two diverging correlation lengths, all quantities will depend on
the ratios $L / \xi$ and $L_{\tau} / \xi_{\tau}$. By considering this and
eq. (\ref{scalexi}) we get
\be 
\rho(0) = \xi^{-(d+z+2)}P(L/\xi,L_{\tau} / \xi_{\tau}) \; ,
\ee
which can be expressed as
\be
\rho(0) = \frac{1}{L^{d+z-2}} \tilde{\rho}(L^{1/ \nu}\delta,L_{\tau}/L^z)
\; ,
\label{finitesizerho}
\ee
where $P$ and $\tilde{\rho}$ are scaling functions. Thus we see that by
keeping the {\em aspect ratio} $a=L_{\tau} / L^z$ constant, the scaling 
function $\tilde{\rho}$ just depends on one argument. This enables us
to locate the critical point, since at criticality $\delta=0$ and
$L^{d+z-2}\rho(0)$ then is independent of $L$, which means that a
plot of $L^{d+z-2}\rho(0)$ vs. $K$ for different system sizes with
constant aspect ratio yields an intersection point at the critical point.
\subsection{Compressibility}
In a similar way as for the stiffness, a scaling expression for the 
compressibilty
\be
\kappa = \frac{\partial n}{\partial \mu}
\ee
can be derived. By imposing a twist in the phase of the order parameter in 
the imaginary time direction instead of the space directions, one obtains 
the scaling form
\be
\kappa \sim \xi^{-(d-z)} \; ,
\ee
and the corresponding finite-size scaling form is
\be
\kappa(0) = \frac{1}{L^{d-z}} \tilde{\kappa}\left( L^{1/ \nu}\delta,
                 \frac{L_{\tau}}{L^z} \right)\;.
\ee

Since by definition $\kappa$ measures the fluctuations in the particle
number, which is given by the $J_i^{\tau}$'s in the classical model
(\ref{action}), $\kappa$ can be expressed in terms of the link
variables as \cite{soerensen}
\be
\kappa(0) = \frac{1}{L^d L_{\tau}} [\langle N_b^2 \rangle - 
                                    \langle N_b \rangle^2]_{\rm av} \; ,
\label{linkkappa}
\ee
where $N_b = \sum_{i,\tau'} J^{\tau}_{i,\tau'}$. 

\subsection{Correlation functions}
In addition to $\nu$ and $z$ a third critical exponent $\eta$ can be 
introduced by considering the 
correlation function $C({\bf r,r}',\tau,\tau')$ for creating a boson
at point $({\bf r},\tau)$ and destructing it at $({\bf r}',\tau')$.
In the quantum phase model (\ref{qph}) this correlation function is given by
\be
C({\bf r,r}',\tau,\tau') = [\langle e^{i[{\hat \phi}_{\bf{r}'}(\tau')-
      {\hat \phi}_{{\bf r}}(\tau)]} \rangle ]_{\rm av} \; ,
\ee
and due to the disorder average $C({\bf r,r}',\tau,\tau')$ will be
translationally invariant, i.e. $C({\bf r,r}',\tau,\tau') = 
C({\bf r-r}',\tau-\tau')$.
We will only be interested in the equal time correlation function
$C_x(r)=C(r,0)$ and the time dependent correlation function 
$C_+(\tau)=C(0,\tau)$, which is essentially a Green's function.

The spatial correlation function $C_x(r)$ can be expressed in
terms of the link variables as \cite{soerensen}
\be
C_x(r) = \left[ \left< \prod_{{\bf r} \in \mbox{path}}
   \exp \left\{ - \frac{1}{K} \left( J^x_{({\bf r},\tau)} + \frac{1}{2}
   \right) \right\} \right>  \right]_{\rm av}
\label{spatialcorr}
\ee
where "path" is a straight line connecting two points a distance 
$r$ apart in the $x$-direction with $\tau$ fixed. 
Since the classical action (\ref{action}) is invariant
under the transformation $J_x \rightarrow - J_x$ it follows that
$C_x(r) = C_x(-r)$.

For the time dependent correlation function a similar expression can be
derived. However, due to the lack of general particle-hole symmetry of
Hamiltonian (\ref{qph}), the correlation functions for particles $C_+(\tau)$
and holes $C_-(\tau)$ will generally be different, except for the case when 
$\mu = n/2$ with $n$ integer, since then statistical
particle-hole symmetry is restored. 
For the single site time dependent correlation function one obtains
\be
C_+^i(\tau) = \left< \prod_{\tau' \in \mbox{path}} \exp \left\{ -\frac{1}{K} 
\left( \frac{1}{2}    + J^{\tau}_{(i,\tau')} - \mu_i \right) \right\} 
\right>  \; ,
\label{imagcorr}
\ee
where "path" is a straight line connecting two points with fixed ${\bf r}$ a
distance $\tau$ apart in the (imaginary) time direction. The site and 
disorder averaged correlation function $C_+(\tau)$ is then defined as 
\be
C_+(\tau) = [C_+^i(\tau)]_{\rm av}.
\label{imagcorrav}
\ee
A similar expression can be derived 
for $C_-(\tau)$. The single site function $C_+^i(\tau)$ corresponds
to the local imaginary time Green's function 
$\langle \Phi_i(\tau)\Phi^+_i(0)\rangle$ in the original Bose Hubbard
model (\ref{bh}).

According to Ref. \cite{fwgf} a scaling form for $C({\bf r},\tau)$ can be  
derived under the assumption that the long-distance and long-time behaviour 
depends on the ratios $r / \xi$ and $\tau / \xi_{\tau}$.
This yields
\be
C({\bf r},\tau) = r^{-(d+z-2+\eta)}f(r/\xi,\tau / \xi^z) \; ,
\ee
with a scaling function $f$ and critical exponent $\eta$. It follows that
at criticality 
\be
C_x(r) \sim r^{-y_x} \quad \mbox{and} \quad C_+(\tau) \sim \tau^{-y_\tau} \;,
\ee
where $y_x$ and $y_\tau$ are given by
\be
y_x = d-2+z+\eta
\label{y_x}
\ee
and 
\be
y_\tau =(d-2+z+\eta) /z \; .
\label{y_tau}
\ee
Thus if one is able to determine $y_x$ and $y_\tau$ in the simulations,
$z$ and $\eta$ can be calculated from 
\be
z=y_x / y_\tau
\label{z}
\ee
and 
\be
\eta = y_x -d + 2 -y_x / y_\tau \;.
\label{eta}
\ee
\section{Monte Carlo Algorithm}
\label{secmc}
The Monte Carlo algorithm has to account for the zero divergence
condition of the link variables. Following Ref. \cite{soerensen} 
we choose our update procedure to consist of local and global moves.
In a local move four link variables
on an elementary plaquette are changed simultaneously. Two of the link 
variables are incremented and the two other are 
decremented by one, thus fulfilling the zero divergence condition.
The inverse of such a local move is also used, i.e. a configuration
where the pluses and minuses are exchanged.
Global moves consist of changing a whole line of link variables extending
all through the system by $\pm 1$ and are performed in all three directions
$x,y$ and $\tau$. We apply periodic boundary conditions, so the zero
divergence condition is also fulfilled. A global move in the $\tau$ 
direction amounts to injecting or destructing a boson, since the $\tau$ 
component $J_i^{\tau}$ of the link variables measures the particle number.
Thus when one explicitly wishes to keep the number of particles constant as 
in Sec. \ref{secpartratio}, global moves in the $\tau$ direction 
have to be either excluded or they have to be performed pairwise, i.e.
one creates and destructs a boson at two different sites simultaneously.

For a given configuration of the link variables we then perform
a local or global move and calculate the energy difference $\Delta E$
between these two states. We use the heat-bath algorithm,
where the new configuration is accepted with probability
\be
w(\Delta E) = \frac{1}{1+\exp(\Delta E /  K)} \; .
\ee
Time is measured in Monte Carlo steps (MCS), where one step corresponds
to a sweep of local and global moves through the whole lattice.

Since we are interested in equilibrium properties of the system 
equilibration is an important issue. In order to have a criterion
for equilibration we proceed as in Ref. \cite{soerensen} where a method
used for spin glasses \cite{spinglasseq} has been adapted. We consider
two replicas $\alpha$ and $\beta$ of a system with the same realization 
of disorder and calculate the ``Hamming''-distance, which is defined as
\be
h^{\nu}_{\alpha, \beta}(t_0) = \left[ \sum_{t=t_0}^{2t_0} \sum_{(i,\tau)}
         \left[ J^{v,\alpha}_{(i,\tau)}(t) - J^{v,\beta}_{(i,\tau)} (t)
     \right]^2 \right]_{\rm av} \; ,
\ee
where $t_0$ is a suitably chosen equilibration time and $\nu=x,y,\tau$. 
We also calculate the ``Hamming''-distance for one replica given by
\be
h^{\nu}_{\alpha} (t_0) = \left[ \sum_{t=t_0}^{2t_0} \sum_{(i,\tau)} 
       \left[ J^{\nu,\alpha}_{(i,\tau)}(t+t_0)   - J^{\nu,\alpha}_{(i,\tau)}
      (t_0)\right]^2 \right]_{\rm av} \; .
\ee
When $t_0$ is chosen sufficiently large, the system is equilibrated and one 
should have 
$h^{\nu}_{\alpha,\beta}(t_0)$ = $h^{\nu}_{\alpha}(t_0)$.
Thus by choosing a sequence of increasing values of $t_0=10,30,100,300,\ldots$
one can give an estimate for the equilibration time from the coincidence of
two values. 

The equilibration time increases drastically with system size.
We found that for smaller systems (e.g. 6x6x9) $\approx 5000$ 
MCS were sufficient, 
whereas the larger lattice sizes (10x10x25) took about $10^5$ MCS. The
equilibration time mainly depends on the system size $L$ in the spatial 
directions, since the disorder, which only enters the transition 
probability of global moves in the $\tau$ direction,
enhances the acceptance rate for this type of moves.
This made it possible to study systems with large $L_{\tau}$ such as 
8x8x200, which became necessary to avoid finite-size effects when 
calculating the probability distribution of the local susceptibility
in Sec. \ref{secbg}.

One also has to consider the disorder average $[\ldots]_{\rm av}$. The number
of samples necessary to get decent statistics strongly depends on
the quantity of interest. For the stiffness (see Section \ref{delta=0.2})
some 100 samples were sufficient while the compressibility has much larger
sample to sample fluctuations and made it necessary to average over
up to 3000 samples. This number of samples was also needed for
the probability distribution of the susceptibility.

The simulations were performed on two massively parallel computers, 
a Parsytech GCel Transputer Cluster with 1024 T-805 transputer nodes and 
a Paragon X/PS 10 with 140 i860 processors.

\section{Bose glass phase} 
\label{secbg}
In this section we systematically investigate the Bose glass phase
and work out the similarities to the quantum Griffiths phase observed
in random transverse Ising systems. We choose
the ''strongest'' possible disorder $\Delta=0.5$, so that
the Mott lobes totally disappear and only the Bose glass-
and the superfluid phase remain. Furthermore throughout this section we 
assume $\mu=1/2$, which corresponds to half filling.
\subsection{Green's functions and the local susceptibility}
In Ref. \cite{fwgf} it is argued that due to a non-vanishing density of 
states at zero energy the disorder averaged imaginary time Green's function 
$C_+(\tau)$ 
decays algebraically
\be
C_+(\tau) \sim \tau^{-1} \; 
\label{greenalg}
\ee
throughout the Bose glass phase.
As a consequence, the superfluid susceptibility $\chi = \int C(\tau) d\tau$
is supposed to diverge in the BG phase and not only a criticality. 
This behaviour is similar to the quantum Griffiths phase observed in other 
models, such as the random transverse Ising chain \cite{young} or the 
Ising spin glass in a transverse magnetic field \cite{rieger,guo}, where 
various susceptibilities also diverge within part of the disordered phase.

To verify the above scenario we calculate the imaginary time Green's
function and the probability distribution of the local susceptibility
in the simulations.
\begin{figure}
\psfig{file=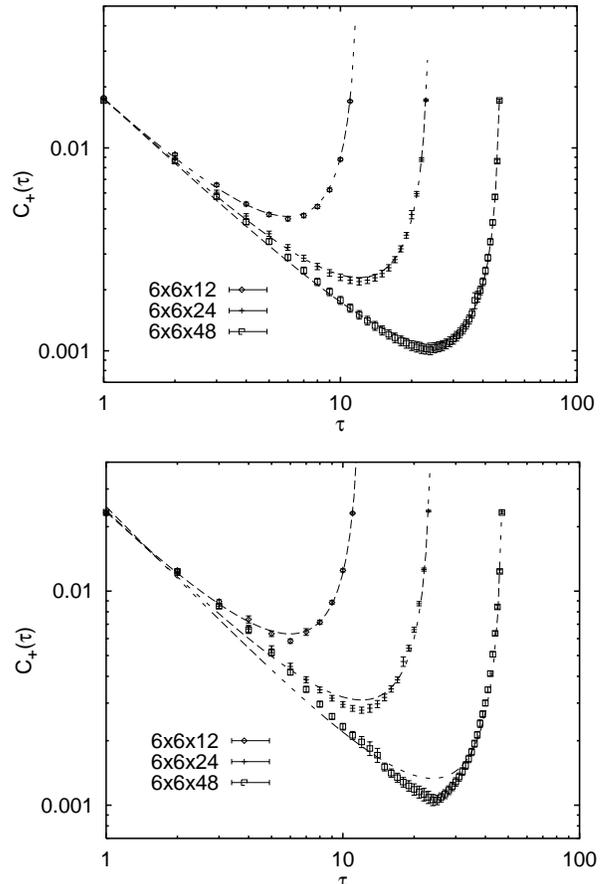,width=\columnwidth}
\caption{The imaginary time Green's function $C_+(\tau)$ for $K=0.17$ (top) 
and $K=0.22$ (bottom) in the Bose glass phase for different system sizes.
The dashed lines are fits to
$C_+(\tau)=a(\tau^{y_\tau}+(L_\tau-\tau)^{y_\tau})$. The exponent $y_\tau$ is 
independent of the system size and independent of $K$. We get 
$y_\tau=-1.08\pm0.03$.}
\label{figgreenbg}
\end{figure}

For the chosen disorder strength $\Delta=0.5$, the phase diagram of Hamiltonian
(\ref{qph}) consists of a Bose glass phase for small hopping amplitudes
$t$ (i.e. small values of the coupling $K$ in the classical model 
(\ref{action})) and a superfluid phase for large values of $t$ (see Fig.
\ref{figphasestrong}). In order
to study the Bose glass phase, we have to locate the critical point
for the BG-SF transition (Point \#1 in Fig. \ref{figphasestrong}) for the 
chemical potential $\mu=0.5$ we are
considering. This can be accomplished by a finite-size
scaling analysis of the stiffness as pointed out in 
Sec. \ref{scalingstiffness}. We find the critical coupling 
$K_C(\Delta=0.5, \mu=0.5)=0.247\pm0.003$. This phase transition was also 
intensively investigated in Ref. \cite{soerensen} in the same representation, 
and the value of $K_C$ reported there perfectly agrees with the one
we found.

We now investigate the BG phase for couplings $K<K_C=0.247$. We start
by calculating the Green's function $C_+(\tau)$ given by eq. 
(\ref{imagcorrav}), which is shown in Fig. \ref{figgreenbg} for couplings 
$K=0.17$ and $K=0.22$.

One observes that $C_+(\tau)$ indeed decays algebraically as
\be
C_+(\tau) \sim \tau^{-1}
\label{greendecay}
\ee
within the error bars and that there is no dependence of the exponent on 
the system size. Similar plots can be made for other values of the coupling 
$K$ in the BG phase, and one obtains the same exponent.

Relation (\ref{greendecay}) leads to a diverging superfluid susceptibility
$\chi \sim \int C(\tau) \, d\tau$. A convenient quantity to further 
characterize the susceptibilty is the probability
distribution of the {\em local}, i.e. single site susceptibility. 
For a diverging superfluid susceptibility a broad distribution of 
local susceptibilities is expected. 
We calculate the local susceptibility $\chi_i$ and the 
corresponding probability distribution 
$P(\chi_i)$, where $\chi_i$ is defined as
\be 
\chi_i = \sum_{\tau=1}^{L_\tau} C_+^i(\tau) \; .
\ee
Since we expect the distribution $P(\chi_i)$ to be very broad, it is
convenient to work on a logarithmic scale, i.e. we consider $P(\ln \chi_i)$
rather than $P(\chi_i)$.
\begin{figure}
\psfig{file=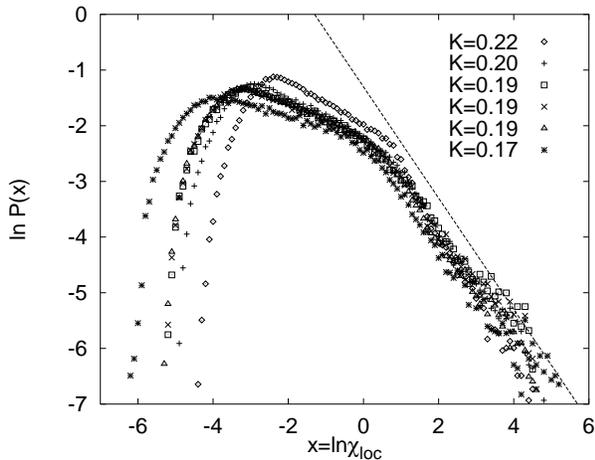,width=\columnwidth}
\caption{The probability distribution $P(\ln \chi_{loc})$ of the
local susceptibility for various values of $K<K_C=0.247$ in the Bose glass 
phase ($\Delta=0.5, \mu=0.5$). The system size is $L=6$ 
and $L_\tau=200$. For $K=0.19$, also data for $L=4$ and $L=10$ are shown, 
which is indistinguishable from $L=6$. The dashed line has slope $-1$.}
\label{figchilocalbg}
\end{figure}

Fig. \ref{figchilocalbg} shows $P(\ln \chi_i)$ for different
couplings $K$ and system sizes in the Bose glass phase. 
One clearly observes asymptotically ($\chi \gg $ 1) a linear relation
between $\ln P(\ln \chi)$ and $\ln \chi$ corresponding to an algebraic decay
$P(\ln \chi) \sim \chi^{-\zeta}$ with $\zeta=1$. Since 
$P(\ln \chi) = \chi P(\chi)$ this is equivalent to
\be
P(\chi) \sim \chi^{-2} \; .
\label{decaychi}
\ee

Following the argument put forward in
Ref. \cite{rieger} for the corresponding quantity in the quantum Griffiths 
phase of the transverse Ising spin glass one can thus introduce a dynamical
exponent $z$:
The large values of $\chi_i$ originate in local clusters containing the site
$i$ that would have a small energy gap if isolated from the rest of the
sample. In the extreme case of vanishing coupling $K$ they simply come from
the sites with a local chemical potential that lends a small energy difference
between single or double occupied sites $n_i=1$ or 2, i.e. $v_i=0$. Thus
the probability for such a value of $\chi_i$ is proportional to the volume of
the system $P(\chi) \sim L^d$. On the other hand, one can relate the length 
scale $L$ with the inverse of the energy scale $\chi$ in the usual way
introducing a dynamical exponent $z$ via
\be                       
P(\chi) \sim \tilde{p}(L^z/\chi) \; .
\ee
With 
\be
\chi P(\chi) \sim L^d \chi^{-\zeta} = (L^z/\chi)^{d/z}
\ee
we obtain 
\be
\ln P(\ln \chi) = -\frac{d}{z} \ln \chi + \mbox{const.}
\label{decayprobchi}
\ee
with $z=d/\zeta$. With $\zeta=1$ it follows that $z=d=2$ throughout the BG 
phase. Note that here $z=d$ in the Bose 
glass phase {\em and} at the critical point, although the two exponents have
their origin in different physics: the dynamical exponent $z$ used in 
eq. (\ref{decayprobchi}) has its origin in purely local effects, whereas
the critical exponent $z$ describes global, collective excitations at 
the critical point. 
\begin{figure}
\psfig{file=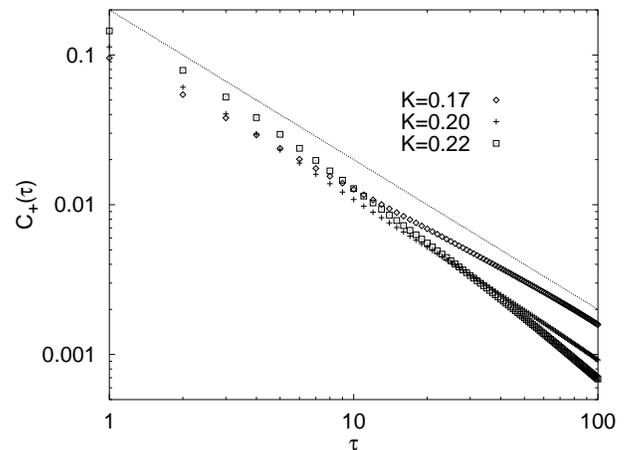,width=\columnwidth}
\caption{$C_+(\tau)$ obtained by Laplace transformation (eq. (\ref{laplace}))
of $P(\chi)$ for different couplings $K$ in the BG phase. The dashed line 
is $\sim \tau^{-1}$. The exponent $-1$ is in agreement with the one 
calculated directly from the correlation function $C_+(\tau)$ 
(see Fig. \ref{figgreenbg}).}
\label{figlaplace}
\end{figure}

If we suppose an exponential decay $C_+^i \sim \exp(-\tau / \chi_i)$, where
$\chi_i$ is an effective inverse correlation length in the $\tau$ direction, 
we can calculate the averaged correlation function 
$C_+(\tau) = [C_+^i (\tau)]_{\rm av}$ from
\be
C_+(\tau) = \int_0^{\infty} P(\chi) \exp(-\tau / \chi) d\chi \; .
\label{laplace}
\ee
Thus we have a nice consistency check for our data by recalculating 
$C_+(\tau)$ from $P(\chi)$. Fig. \ref{figlaplace} shows the resulting
curves for $C_+(\tau)$ for different values of $K<K_C=0.247$. By comparison 
with the dotted line, one again establishes that $C_+(\tau) \sim \tau^{-1}$.

The asymptotic decay of $P(\chi) \sim \chi^{-2}$ that can be clearly seen
in Fig. \ref{figchilocalbg} indicates that the underlying physics  is
purely local. In order to see this
we consider the imaginary time Green's function for $T=0$ in the limit 
of zero hopping $t=0$. Starting from the definition
\be
C_+^i(\tau) = \langle n_i^0 | \Phi_i(\tau) \Phi_i^+(0)| n_i^0 \rangle
\ee
and substituting $\Phi(\tau) = e^{\tau H} \Phi(0) e^{-\tau H}$,
one obtains
\begin{equation}
C_+^i(\tau) \sim e^{\tau[\epsilon(n_i^0) - \epsilon(n_i^0+1)]}\; ,
\end{equation}
where $n_i^0$ is the ground state occupation number at site $i$.
From Hamiltonian (\ref{qph}) the energy difference for the states with
$n_i$ and $n_i+1$ bosons at site $i$  for $t=0$ can be readily 
calculated. We obtain 
$\epsilon(n_i) - \epsilon(n_i+1) = -(n_i+1/2 -\mu -v_i)$. Since we are
working at half filling $\mu = 1/2$, and $v_i \in [-1/2,1/2]$, 
it follows that the on-site energy is minimized by $n_i=0$ for $v_i < 0$ and 
$n_i=1$ for $v_i >0$. For $v_i=0$ the states with $n_i=0$ and $n_i=1$
are degenerate.
As a result we obtain
\be
C_+^i(\tau) = 
\left\{ 
\begin{array}{rr} e^{v_i\tau} & v_i < 0\\
e^{(v_i-1)\tau} & v_i > 0
\end{array} 
\right. \; .
\ee
From the definition $\chi_i(\omega=0) = \int_0^{\infty} d\tau C_+^i(\tau)$
it then follows
\be
\chi_i(\omega=0) = \left\{ \begin{array}{rl} \frac{1}{|v_i|} & v_i <0\\
\frac{1}{1-v_i} & v_i >0 \end{array} \right. \;,          
\label{chi}
\ee
and since $P(v)$  is uniform in the interval $[-1/2,1/2]$, one finally
obtains 
\be
P(\chi) \sim \frac{1}{\chi^2} \; .
\label{decaychiana}
\ee
From this we see that the observed decay (\ref{decaychi}) of $P(\chi)$
can be explained under the assumption that the excitations in the BG 
phase are fully {\em localized} and the sites are essentially decoupled.
In particular it turns out that increasing the coupling $K$ between sites does 
not change the strength of the singularity in (\ref{decaychi}) significantly,
implying that the dynamical exponent $z$ does not change from $K=0$ (z=2)
to $K_C$. This is in contrast to what happens in the otherwise equivalent
quantum Griffiths phase in random transverse Ising models \cite{young,rieger},
where an increase of the coupling between the spins leads to a continous
change of the strength of the Griffiths singularities, i.e. to a 
continously varying dynamical exponent.

The reason for this different behaviour of the Bose glass phase in the
disordered boson Hubbard model and the Griffiths phase in the random
transverse Ising model is the fact that the former is related to an XY-model
whereas the latter is an Ising model. The effect of strongly coupled clusters
on the dynamics (in imaginary time) is much weaker for vector spins
than for Ising spins in analogy to the relaxational dynamics in classical
spin models within the Griffiths phase \cite{brayrandeira}.

To clarify this point and to obtain information about the localization of 
the excitations in the present model, we consider the participation ratio in 
the following section.

\subsection{Participation ratio}
\label{secpartratio}
In the context of localization in disordered quantum mechanical systems
one usually studies a quantity called the participation ratio, which is
defined with respect to one-particle excitations. Given the latter in the
space representation (e.g. $\psi_i$) one obtains via the local probabilities
$\rho_i = |\psi_i|^2$ the information about the degree of localization
of this one-particle excitation by considering the ratio of moments.
Since the MC algorithm we use provides us with information on
the ground state exclusively (not on excitations), we devise
an alternative method: we add an extra particle to the system and study how
it distributes its local probabilities $\rho_i$ among the lattice sites $i$.
To this end we have to keep track of the local particle densities of the
system with and without this extra boson - the difference is then interpreted
as the probability density $\rho_i$ of the extra boson. The degree of
localization can then be read from the participation ratio
\be
p_L = \left[ \sum_{i=1}^N \langle \rho_i \rangle^2 \right]_{\rm av}^{-1} \; .
\label{partratio1}
\ee
For a fully localized state $\rho_i = \delta_{ij}$ it is 
$p_L = {\cal O} (1)$, whereas for a completely delocalized state 
$\rho_i = 1/N$ one has $p_L = {\cal O}(N)$.  
\begin{figure}
\psfig{file=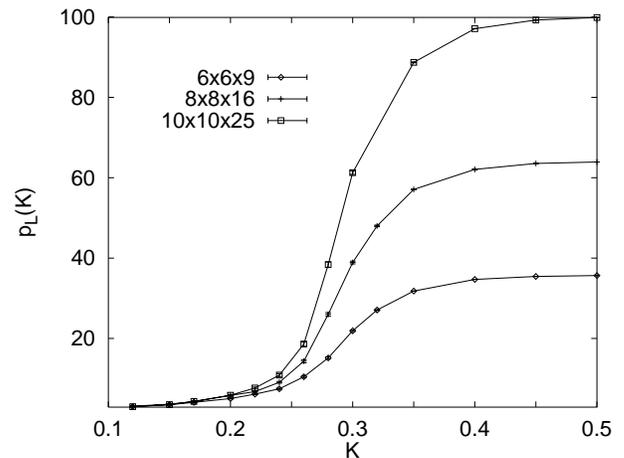,width=\columnwidth}
\caption{The participation ratio $p_L(K)$ defined in eq. (\ref{partratio1})
as a function of $K$ for various system sizes $(\Delta=0.5, \mu=0.5)$. 
Within the Bose glass phase ($K<K_C=0.247$) $p_L$ approaches a constant.}
\label{figpartratio1}
\end{figure}
To calculate $p_L$ in the simulations we use a replica method. We
consider two replicas $\alpha$ and $\beta$ with the same realization
of disorder and with fixed number of particles. Replica $\alpha$ is 
initialized with $N/2+1$ and
replica $\beta$ with $N/2$ bosons, corresponding to the case of half filling
$\mu=1/2$ considered in the preceding calculations. We then calculate
the probability $\rho_i$ of finding the excess boson at site $i$ by
\be
\rho_i = \frac{1}{L_\tau} \sum_{\tau'=1}^{L_\tau} (J_{i,\tau'}^{\tau,\alpha} 
         - J_{i,\tau'}^{\tau,\beta}) \; .
\ee

The restriction of working with a fixed number of particles in each
replica makes it necessary to specifically consider the global moves
in the $\tau$ direction in the Monte Carlo algorithm, since they
amount to either injecting or destructing a boson. In a first attempt one 
can simply eliminate this type of move, but it turned out that it was 
impossible to equilibrate the systems within a reasonable amount of
computer time this way. Alternatively, one can also keep the number of
particles constant by injecting and destructing a particle at two different 
sites simultaneously. We found that by randomly choosing two sites where at
one site a particle is injected and at the other a particle is destructed,
the equilibration time is drastically reduced, enabling us to obtain
equilibrium results.    
\begin{figure}
\psfig{file=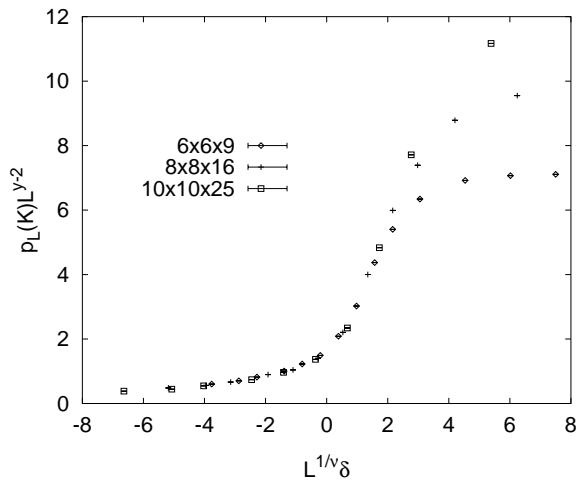,width=\columnwidth}
\caption{Scaling Plot of the participation ratio $p_L(K)$ with $\nu=0.9$ 
and $y=1.0\pm 0.1$.}
\label{figpartscale1}
\end{figure}
Fig. \ref{figpartratio1} shows $p_L$ for different system
sizes with fixed aspect ratio $1/4$ and for different couplings $K$.
One observes that the participation ratio increases with $K$ until
it eventually saturates. At saturation it is indeed as
expected $p = {\cal O}(N)$,
which tells us that in the superfluid phase the extra boson is 
completely delocalized.
For lower couplings however, the participation ratio approaches
1 with decreasing $K$ independent of the system size, 
which indicates that the extra boson is localized on a single site. 
We remark again that the phase
transition from the Bose glass to the superfluid phase occurs at $K_C=0.247$.
 
By inspection of Fig. \ref{figpartratio1} and motivated by a recent result
within mean-field theory \cite{pazmandipratio}, one might expect that the 
participation ratio satisfies a scaling law. As an ansatz for a 
scaling relation we make
\be
\frac{p_L(K)}{L^2} = L^{-y}\tilde{q}(L^{1/\nu}\delta ,L_{\tau}/L^z) \;,
\label{partscale}
\ee
where the scaling function $\tilde{q}$ and an additional exponent $y$ 
have been introduced. $\delta=(K-K_C)/K_C$ is the distance from the critical
point and $\nu$ the correlation length exponent. The values $K_C=0.247$ 
and $\nu=0.9$ can be obtained from a scaling analysis of the 
stiffness according to Sec. \ref{scalingstiffness} and the values found are in 
agreement with the values reported in Ref. \cite{soerensen}.
We emphasize that we work at fixed aspect ratio $L_{\tau}/L^z=1/4$,
so that the scaling function $\tilde{q}$ just depends on one argument and
the only variable to be determined is $y$.
Fig. \ref{figpartscale1} shows a scaling plot of the
participation ratio. The data scales nicely as long as the participation
ratio is not saturated. We obtain best scaling for $y=1.0 \pm 0.1$.
\begin{figure}
\psfig{file=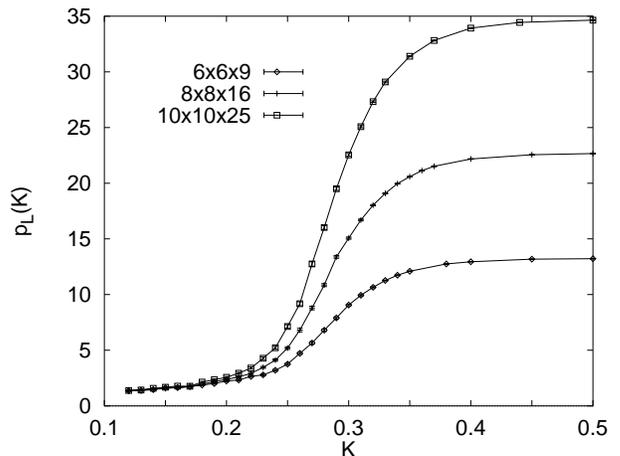,width=\columnwidth}
\caption{The participation ratio $\tilde{p}_L(K)$ defined in eq.
(\ref{partratio2}) as a function of $K$ for various system sizes. Within 
the Bose glass phase ($K<K_C=0.247$) $\tilde{p}_L$ approaches a constant.}
\label{figpartratio2}
\end{figure}
We also considered different moments of the probability $\rho_i$. For 
instance, we calculated
\be
\tilde{p}_L= \left[ \left\langle \frac{(\sum_i \rho_i^2)^2}{\sum_i \rho_i^4}
     \right\rangle \right]_{\rm av} \; ,
\label{partratio2}
\ee
which is shown in Fig. \ref{figpartratio2}. One observes that the curves
look very similar to the ones in Fig. \ref{figpartratio1} and one obtains
the same information about the localization of the excitations: 
for $K<K_C=0.247$ $\tilde{p}_L$ approaches 1 independent of the system size 
indicating again the localization of the excitations in the BG phase,
whereas for larger values of $K$ it is $\tilde{p}_L \sim {\cal O}(N)$
meaning that the excitations are delocalized. We found that it is easier
to calculate $\tilde{p}_L$ in the simulations than $p_L(K)$, as it turned
out that the equilibration time for $\tilde{p}_L$ is much shorter than
for $p_L$. This is the reason why we have more data points for
$\tilde{p}_L$, making the curves look smoother.
A scaling plot of $\tilde{p}_L$ according to eq. (\ref{partscale}) is shown 
in Fig. \ref{figpartscale2}. We find the same exponent $y=1.0 \pm 0.1$ as
for the other definition of the participation ratio in eq. (\ref{partratio1}).
\begin{figure}
\psfig{file=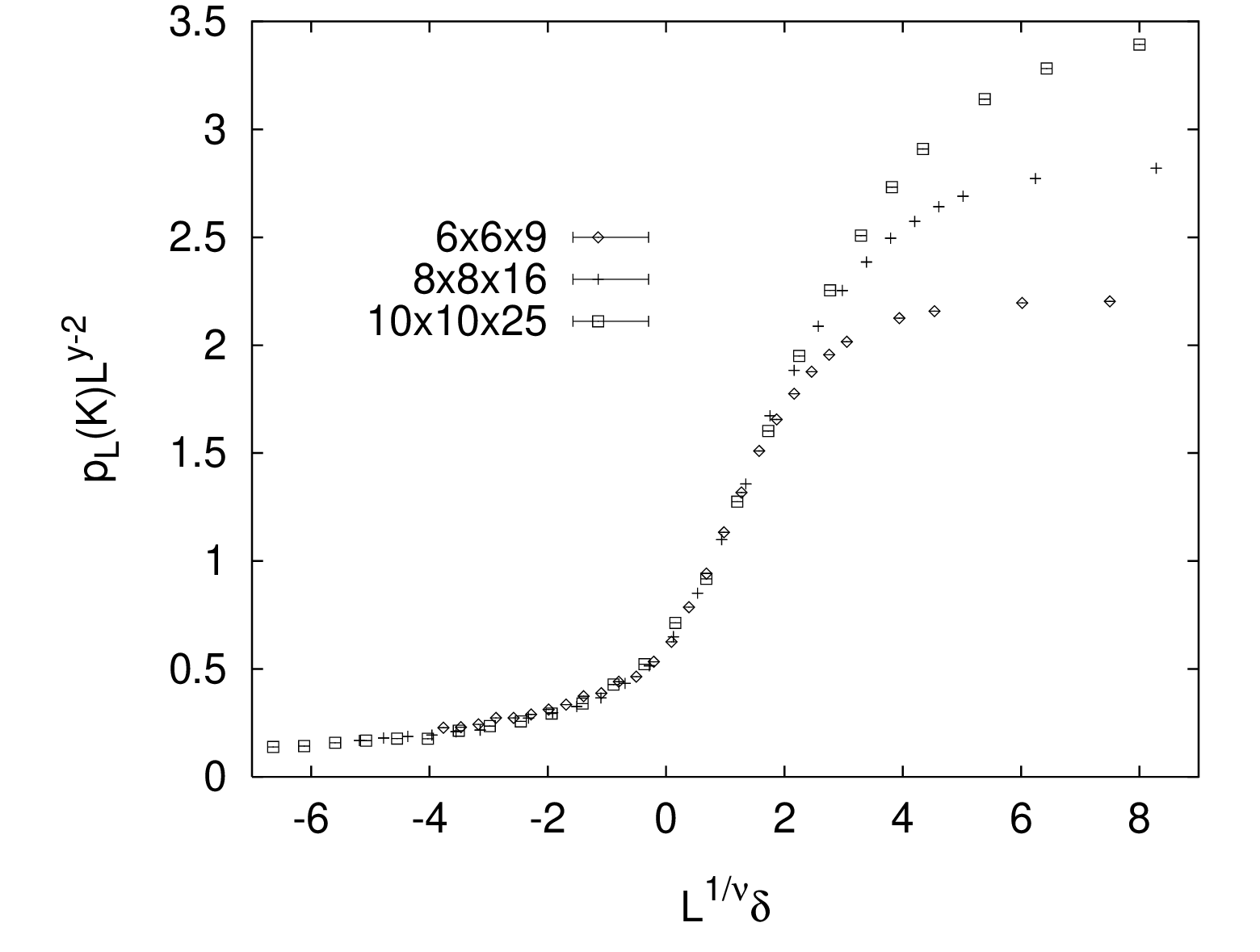,width=\columnwidth}
\caption{Scaling Plot of the participation ratio $\tilde{p}_L(K)$ with 
$\nu=0.9$ and $y=1.0\pm0.1$.}
\label{figpartscale2}
\end{figure}
\section{Phase diagram and phase transitions for weak disorder}
\label{secweakdis}
\label{delta=0.2}

In this section we consider the model (\ref{action}) for disorder
strength $\Delta=0.2$. In particular we examine the phase diagram at the
tip of the lobe, i.e. at $\mu=1$ (Point \#1 in Fig. \ref{figphaseweak}), 
where it is not clear whether 
there is a direct Mott insulator to superconductor phase transition or
if an intervening Bose glass phase exists. Without disorder this phase
transition is in the universality class of the 3D XY-model.
In their scaling theory Fisher {\em et al.} argue that for
every disorder strength $\Delta\neq 0$ the Mott lobes are surrounded
by a Bose glass phase even at the tip of the lobe and consequently,
the insulator-superconductor transition always occurs from the Bose glass
phase. To check this prediction and to determine the critical exponents
we perform a finite-size scaling analysis of the stiffness and the
compressibility and also investigate the Mott insulator to Bose glass
transition. Furthermore we calculate the probability distribution 
of the local susceptibility $P(\ln \chi)$
in the Mott insulating and the superfluid phase.

\subsection{Finite-size scaling analysis at the tip of the lobe$(\mu=1)$}
The relevant quantities of interest are the stiffness $\rho$ and the 
compressibility $\kappa$, since they allow for a distinction of the different 
phases. We remark again that the stiffness vanishes in the Mott insulating 
and the Bose glass phase but is finite in the superfluid phase, whereas
the compressibility is zero in the Mott insulating phase and finite in
the Bose glass and the superfluid phase. 

Since for a finite-size scaling analysis according to Sec. 
\ref{secscalingth} the aspect ratio $L_{\tau} / L^z$ has to be kept 
constant, and the dynamical exponent $z$
is not known a priori, we first have to make a choice for $z$.
However, by calculating the correlation functions 
(\ref{spatialcorr}) and (\ref{imagcorrav}) at criticality, we obtain $z$ 
independent of the aspect ratio, and thus we have a consistency check 
for the assumed value of $z$.
\begin{figure}
\psfig{file=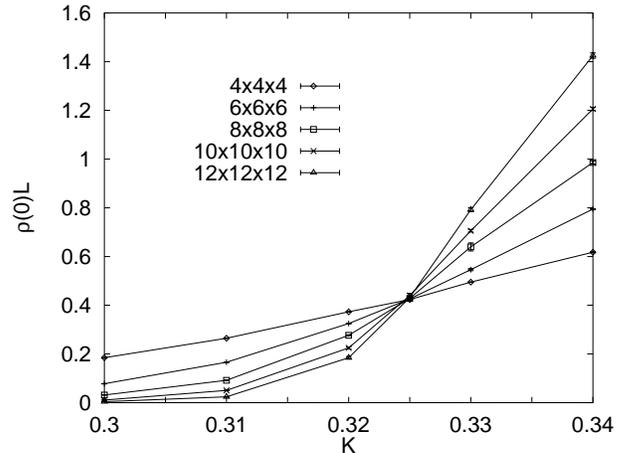,width=\columnwidth}
\caption{The stiffness $\rho(0)L$ at the tip of the lobe ($\Delta=0.2,
\mu = 1.0$). The aspect ratio is constant for $z=1$. The intersection
point is seen to be at $K_C(\Delta=0.2,\mu=1)=0.325\pm 0.003$.}
\label{figstiffness}
\end{figure}
We proceed by first assuming $z=1$ which is the value for the pure case and
remark that the following calculation of the correlation functions shows that
this is indeed the correct value. 
We calculate the stiffness $\rho$ as  given by eq. (\ref{linkstiffness}) and 
perform a finite-size scaling plot
guided by eq. (\ref{finitesizerho}), i.e. we plot $L^{d+z-2}\rho$ vs. $K$ for
different system sizes but with constant aspect ratio $L_\tau/L^z=1$. 
At the critical point
$L\rho$ should be independent of $L$, so at criticality we expect
an intersection of the different curves. Fig. \ref{figstiffness}
shows the corresponding plot. One clearly observes an intersection point
at the critical coupling $K_C(\Delta=0.2, \mu=1) =0.325 \pm 0.002$. 
For $K \le K_C$ the stiffness scales to zero with increasing system 
size which indicates an insulating phase.
According to Ref. \cite{fwgf}, the Bose glass to superfluid
transition always has  $z=2$, which suggests that the transition
from the insulating side is from the Mott insulator rather than 
the Bose glass phase. The critical point is marked as point \# 1 in Fig. 
\ref{figphaseweak}.

The critical exponent $\nu$ can be obtained from a scaling plot
of the stiffness. In Fig. \ref{figscalestiff} we plot 
$L^{d+z-2}\rho$ vs. $L^{1/ \nu}\delta$ and adjust
$\nu$ as to achieve best data collapse where $\delta=(K-K_C)/K_C$ with 
$K_C=0.325$. One observes that the data scale nicely and we obtain
$\nu = 0.7 \pm 0.1$.

\begin{figure}
\psfig{file=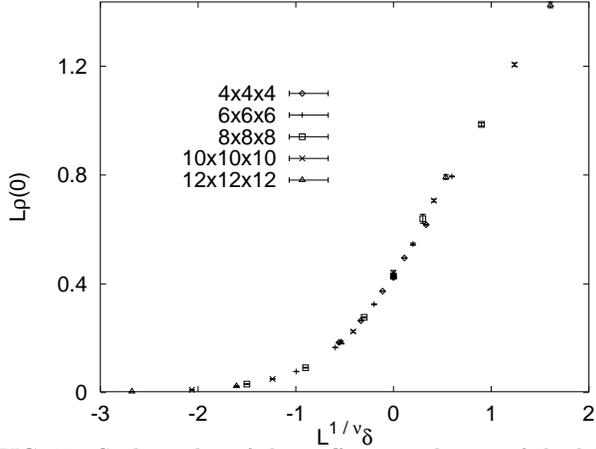,width=\columnwidth}
\caption{Scaling plot of the stiffness at the tip of the lobe 
$(\Delta=0.2, \mu=1)$ with $K_C=0.325$ and $\nu = 0.7 \pm 0.1$.}
\label{figscalestiff}
\end{figure}

We proceed by  calculating the compressibility $\kappa$ according to eq. 
(\ref{linkkappa}). Fig. \ref{figcompress} shows a finite-size scaling plot 
of $\kappa$ with $z=1$. One again observes an intersection point close to 
$K=0.325$, even though the intersection point is not as good as the one for 
the stiffness. This is due to fact that the compressibility has much
larger sample to sample fluctuations than the stiffness and so it is
more difficult to obtain good statistics. Note that the number of samples
for the compressibility already is 2500, whereas for the stiffness
250 samples were sufficient. The large fluctuations also limited
the range of system sizes we were able to simulate for the compressibility,
so we only have data up to systems of size $8\times8\times8$.
For couplings $K < K_C=0.325$ we see
that $L\kappa$ decreases with increasing system size, which 
indicates that in the thermodynamic limit $\kappa=0$.
\begin{figure}
\psfig{file=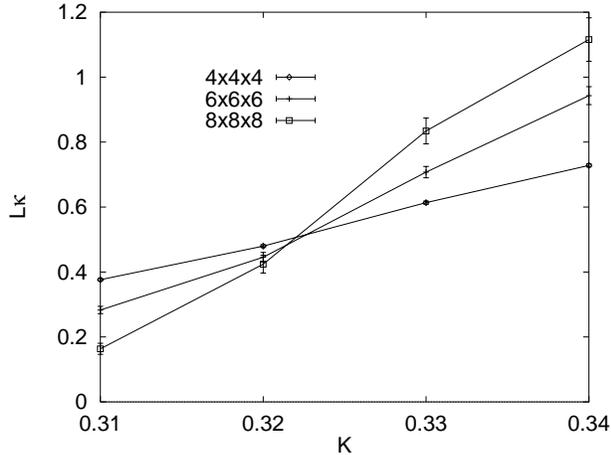,width=\columnwidth}
\caption{The compressibilty at the tip of the lobe ($\Delta=0.2, \mu=1.0)$
for different sytem sizes. The aspect ratio is constant for $z=1$. The
intersection point coincides with the one for the stiffness in Fig.
\ref{figstiffness} within the error bars.}
\label{figcompress}
\end{figure}
Considering the above results for the stiffness and the compressibility,
we conclude that below $K_C=0.325$ we have indeed a Mott insulating phase and 
for $K>K_C$ a superconducting phase, in particular there is no sign of
an intervening Bose glass phase.
\begin{figure}
\psfig{file=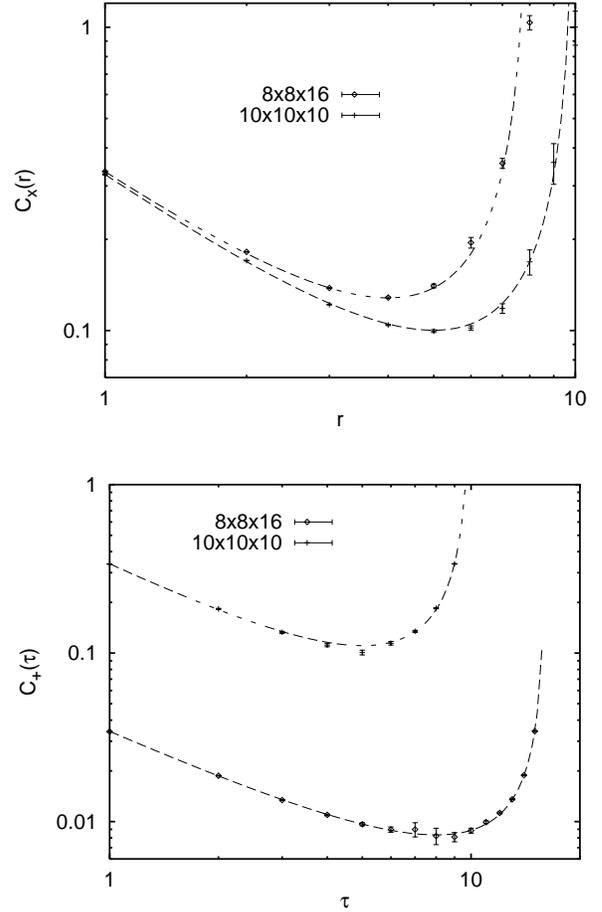,width=\columnwidth}
\caption{The correlation functions $C_x(r)$ (top) and $C_+(\tau)$ (bottom)
for system sizes 8x8x16 and 10x10x10 at the critical point $K_C=0.325$.
The dashed lines are fits to $C_x(r)=a(r^{y_x}+(L-r)^{y_x})$ and 
$C_+(\tau)=b(\tau^{y_{\tau}}+(L_\tau-\tau)^{y_{\tau}})$
respectively. We obtain $z=y_x / y_{\tau} = 1$ independent of the shape 
of the sample.}
\label{figcorr}
\end{figure}
We can now go for the consistency check of our assumed value of $z=1$.
Therefore we measure the spatial- and imaginary time correlation functions
as given by eq. (\ref{spatialcorr}) and (\ref{imagcorrav}). From
these functions we obtain the exponents $y_x$ and $y_{\tau}$
and then $z$ is simply given by eq. (\ref{z}), i.e. $z=y_x / y_{\tau}$. 
Fig. \ref{figcorr} shows the correlation functions for different system 
sizes at criticality $K_C=0.325$. By fitting to the functions 
$C_+(\tau) =  a(\tau^{-y_{\tau}}+(L_{\tau}-\tau)^{-y_{\tau}})$
and $C_x(r) =a'(r^{-y_x}+(L-r)^{-y_x})$ 
respectively we obtain $y_{\tau} = 1.11 \pm 0.02$,
 $y_x = 1.07 \pm 0.02$ for system size 10x10x10 and $y_\tau=1.10 \pm 0.02$, 
$y_x=1.10 \pm 0.02$ for system size 8x8x16.
This yields $z=y_x / y_{\tau} = 1$ in both cases, confirming the value
used in the scaling analysis of the stiffness and the compressibility and
demonstrating that $z$ is indeed independent of the shape of the sample.
Additionally from $y_x$ and $y_\tau$ the critical exponent
$\eta$ can be obtained according to eq. (\ref{eta}), which yields
$\eta = 0.1  \pm 0.1$.

To summarize the above scaling analysis, we find a direct Mott insulator
to superconductor transition with no intervening Bose glass phase
at the multicritical point $\mu =1$. The critical exponents are
$z=1$, $\nu =0.7 \pm 0.1$ and $\eta =0.1 \pm 0.1$. These are very close to 
the exponents obtained for the 3D XY-model both analytically from 
$\epsilon$-expansions \cite{leguillou} $\nu=0.669 \pm 0.002$, 
$\eta=0.033 \pm 0.004$ and by Monte Carlo simulations \cite{soerensen} where
$\nu = 0.667$ and $\eta \approx 0.02$ were obtained. 
Thus we conclude that the universality class of the transition is not 
changed by weak disorder. However, this might be different for stronger 
disorder, and for instance for $\Delta=0.4$ we get an effective exponent
$z=1.4$. 

Furthermore we want to comment on the result $\nu=0.7$. This value
of $\nu$ seems to contradict the inequality $\nu \ge 2/d$
\cite{harris,chayes}, where $d$ is the
dimension in which the system is disordered (i.e. here $d=2$ ), which
should be applicable to a broad class of phase transitions that can be  
triggered by the strength of the disorder.
However, by comparing the critical couplings
for the MI-SF transition for the pure case 
$K_C(\Delta=0,\mu=1)=0.333\pm 0.003$ 
and weak disorder $K_C(\Delta=0.2,\mu=1)=0.325 \pm 0.002$ it becomes evident, 
that the 
position of the tip of the lobe only depends very weakly on the disorder 
strength or is even independent of it, at least for small disorder. 
Thus this transition cannot be triggered by varying the disorder strength
and therefore the inequality possibly does not apply here. We also want
to mention the possibility, that our data are not yet in the asymptotic 
scaling regime and therefore the measured exponent $\nu$ is only 
an effective exponent for small length scales, although we find this very
unlikely due to the quality of the scaling plots.

We remark that we also find an intersection point in the finite size
scaling plot of the stiffness at $K=0.32$ 
(which is close to the critical coupling $K_C=0.325$ we find for $z=1$) 
when assuming
$z=2$, even though this intersection point is not as convincing as the one
for $z=1$. However, as pointed out above, by calculating
the correlation functions we always obtain $z=1$, independent
of the shape of the sample. Furthermore there is no intersection point
in the plot of the compressibility for $z=2$. Thus a transition with $z=2$,
which could indicate a transition from the Bose glass phase, 
is ruled out by the above results.

\subsection{The Mott Insulator - Bose glass transition}
As we have seen the BG phase is quite similar to the quantum Griffith
phase occuring in random transverse Ising models. One can extend the analogy
to the other phases, too, and formulate the following dictionary (see Table
\ref{tabdict}):

One can identify the Mott insulating and paramagnetic phases as the
disordered phases, the Bose glass and the Griffiths phase as the weakly or
locally ordered phases and the superfluid and ferromagnetic phase
as the ordered phases. The transition from the insulating BG phase
to the superconducting phase as well as the transition from the paramagnetic
Griffiths phase to the ferromagnetic phase are both second order, whereas the
MI-BG and PM-Griffiths ``transitions'' are in fact {\em not} real phase 
transitions:
only some susceptibility starts to diverge due to local effects, which for
the Bose glass phase also implies that it becomes compressible.
As in the context of the Griffiths phase, where a continuously increasing
(coming from the disordered phase) dynamical exponent leads to this
divergence, no diverging length scale is connected with the closing of the gap upon entering the Bose glass phase (coming from the Mott insulator).

\begin{figure}
\psfig{file=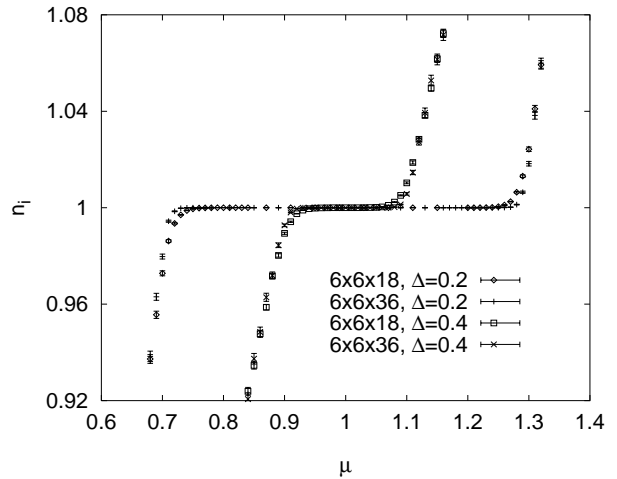,width=\columnwidth}
\caption{Number of particles per site vs. chemical potential at $K=0.19$
for different system sizes and different disorder strengths $\Delta$.
In the Mott phase the particle number is fixed at an integer value. For
increasing disorder $(\Delta=0.4)$ the Mott phase shrinks.} 
\label{figpnumber}
\end{figure}
As a consequence we cannot locate the MI-BG transition via finite size
scaling. To get an idea of the phase diagram,
we measure the particle number $n_i$ per site as a function of the chemical
potential $\mu$ at fixed coupling $K=0.19$. This corresponds to moving
along the line through point \#2 in Fig. \ref{figphaseweak}.
We choose the value of $K=0.19$ as
to be below the critical coupling $K_C(\Delta=0.2,\mu=0.5)=0.20$ for
the BG-SF transition (Point \#3 in Fig. \ref{figphaseweak}).  
Such a plot is shown in Fig. \ref{figpnumber} for $K=0.19$. 
The different curves correspond to different system sizes.
One observes that starting at the chemical potential $\mu=0.65$ the particle 
number per site increases
with roughly constant slope. This means, that the compressibility $\kappa=
\partial n / \partial \mu$ is indeed finite in a certain range of the
chemical potential. For $\mu \approx 0.73$ the particle number per site
reaches one, meaning that we are in the Mott insulating phase. Increasing
the chemical potential further does not change the number of particles, which
means that the compressibility vanishes as expected. For higher values of 
$\mu$, we leave the Mott phase and again have a phase with non vanishing 
compressibility. For stronger disorder ($\Delta=0.4$) one observes a similar
behaviour, but the range of the chemical potential where the particle number 
is constant is smaller. This indicates that with increasing disorder
the Mott lobes indeed shrink, as the disorder reduces the energy gap in
the MI phase.  

In Fig. \ref{figpnumbertip} we show the number of particles vs. chemical
potential near the tip of the lobe. It can be seen that at the critical 
point $K_C(\Delta=0.2, \mu =1)=0.325$ there is no longer a range of the 
chemical potential where the number of particles is constant demonstrating 
that the MI phase disappears at $K_C$.
\begin{figure}
\psfig{file=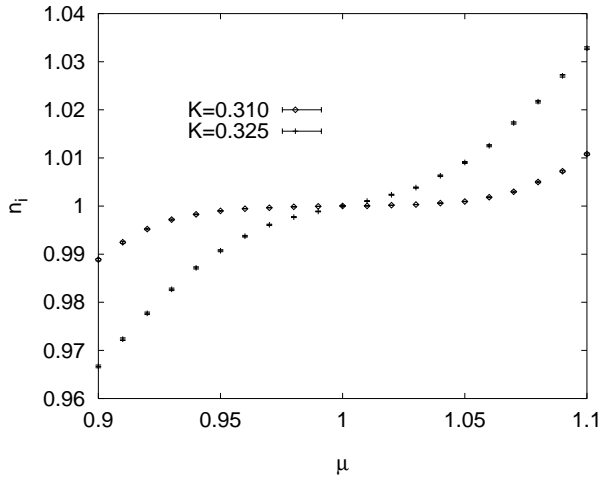,width=\columnwidth}
\caption{Number of particles per site vs. chemical potential for $\Delta=0.2$ 
near the tip of the lobe for system size 6x6x18,
demonstrating that the Mott lobe disappears at $K_C=0.325$.}
\label{figpnumbertip}
\end{figure}
\begin{figure}
\psfig{file=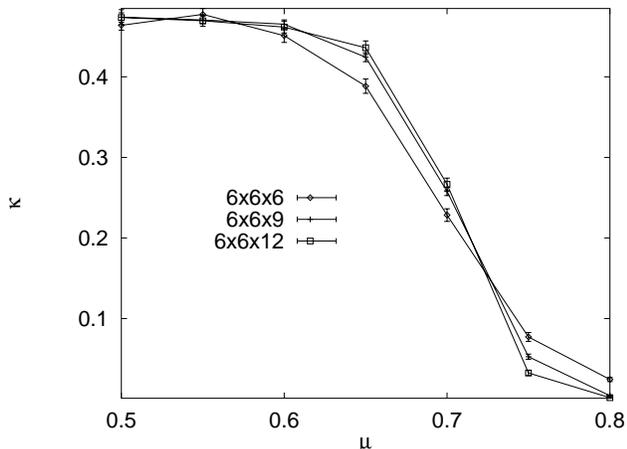,width=\columnwidth}
\caption{The compressibility $\kappa$ as a function of the chemical potential 
$\mu$ for the Mott insulator to Bose glass transition at fixed $K=0.19$
($\Delta=0.2$).}
\label{figcomp022}
\end{figure}
We also calculate the compressibility directly as function of $\mu$ at the 
same value of $K=0.19$ as above, which is shown in Fig. \ref{figcomp022}.
As long as the typical length scale $\xi$ at the MI-BG transition
(which does {\em not} diverge here, as we mentioned above) is smaller than
the system size $L$, the compressibility $\kappa$ will not depend 
significantly on $L$. Indeed we observe a much stronger dependence on
$L_tau$, which is due to the closing of the gap at the MI-BG transition.
In Fig. \ref{figcomp022} we demonstrate this effect by showing data for
increasing system size in the imaginary time direction and constant system
size in the space direction. The curves intersect at $\mu=0.73$, which
corresponds to the left edge of the MI-plateau in Fig. \ref{figpnumber}, i.e.
the MI-BG transition point. For increasing $L_\tau$ the curves become steeper 
and one might guess that in the limit $L_\tau \rightarrow \infty$ the
compressibility jumps discontinously from $\kappa =0$ for $\mu > 0.73$
(in the MI-phase) to a finite value ($\kappa \approx 0.5$) for $\mu < 0.73$
(in the BG phase). Unfortunately the sample to
sample fluctuations of $\kappa$ are extremely large so that we could
not get good enough statistics for larger values of $L_\tau$. 

A discontinuity in the compressibility would support the conjecture
that the MI-BG transition is first-order \cite{freericks}.  At a
generic first order phase transition one expects a phase coexistence,
which in our case means the coexistence of the compressible BG and the
incompressible MI phase at $\mu\approx0.73$.  Such a possibility can
be studied systematically \cite{firstorder} via the calculation of the
probability distribution $P(\kappa)$ of the compressibility in finite
systems. If the MI-BG transition is first order and shows phase
coexistence, one expects a double peak structure in $P(\kappa)$ with a
peak at zero (from MI phase) and at $\approx 0.5$ (from the BG phase).
However, our data for $P(\kappa)$ (see Fig. \ref{figprobcomp})
show only a single peak moving from
$\kappa=0$ to $\kappa\sim0.5$ when increasing $\mu$ from below to
above the transition, which excludes phase coexistence.
\begin{figure}
\psfig{file=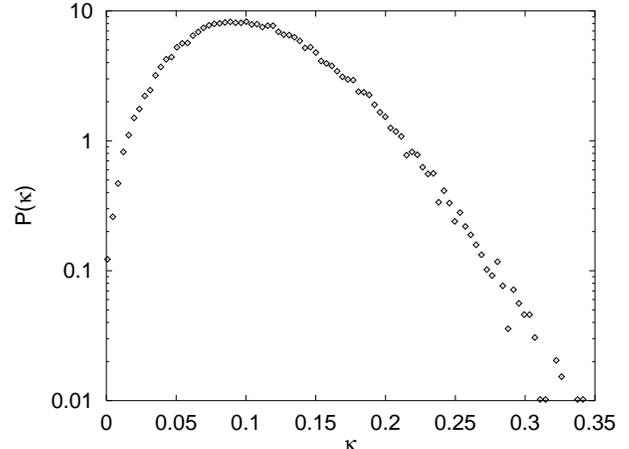,width=\columnwidth}
\caption{The probability distribution $P(\kappa)$ of the compressibility
for system size 6x6x9 at $\mu=0.73$ and $K=0.19$ ($\Delta=0.2$). Note the 
logarithmic scale of the y-axis and the exponential tail of the distribution.}
\label{figprobcomp}
\end{figure}

\subsection{The local susceptibility in the Mott insulating and superfluid
phase}
Finally we consider the probability distribution $P(\ln\chi)$ of the 
local susceptibility in the Mott insulating and the superfluid phase.
We fix the chemical potential at $\mu=1$, i.e. we are at the tip of the lobe 
and calculate $P(\ln \chi)$ for different couplings $K$. The resulting 
curves are shown in Fig. \ref{figchilocalmi}. From the preceding section we
know that the Mott insulator to superfluid transition for $\Delta=0.2$ 
and $\mu=1.0$ is at $K_C=0.325$.
For the smallest value of $K=0.250$, which clearly is in the Mott insulating
phase, one observes that $P(\ln \chi)$ has no tail as in the Bose glass
phase but rather abruptly decays to a very small value. This indicates
that in the MI phase one has indeed a finite energy gap. At criticality,
i.e. at $K=0.325$, the imaginary time Green's function decays algebraically
with an exponent close to one ($\approx 1.1$, see Fig. \ref{figcorr}).
Since $\int d\tau C(\tau) \sim \int d\chi \chi P(\chi)$, 
we again expect a probability distribution
$P(\ln \chi)$ with a $1/ \chi$ tail, which is verified by comparison with
the dashed line in Fig. \ref{figchilocalmi}. For higher values of $K$ in the 
superfluid phase the curves are shifted to the right, and one again has an 
algebraic tail. 
\begin{figure}
\psfig{file=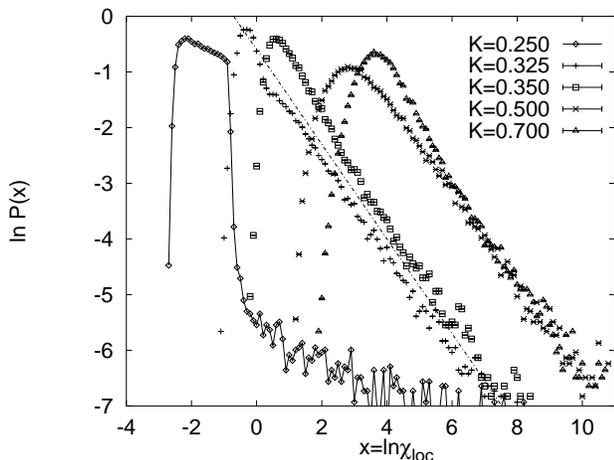,width=\columnwidth}
\caption{The probability distribution $P(\ln \chi_{loc})$ of the local
susceptibility in the Mott insulating and the superfluid phase as a function
of $K$ for $\Delta=0.2$ and $\mu=1$. The MI-SF
transition is at $K_C=0.325$. In the MI phase ($K=0.250$) the distribution
$P(\ln \chi)$ has no tail, indicating a finite energy gap. At criticality
and in the SF phase one recovers a tail. The dashed line 
has slope -0.9.}
\label{figchilocalmi}
\end{figure}

\section{Summary and discussion}
\label{secconclusion}
Concluding we have shown that the Bose glass phase shares many features
with the Griffiths phase in random transverse field Ising systems.
We find an algebraic decay of the imaginary time Green's function,
a broad probability distribution of the local susceptibility with
an $1 / \chi^2$ tail and thus a diverging global superfluid susceptibility.
The occuring singularities can be parametrized by a dynamical exponent $z=2$
which is constant in the Bose glass phase. This dynamical exponent $z$ is
equal to the critical dynamical exponent $z=d=2$ for the Bose glass 
superfluid transition and it is an open question whether this equality 
holds in general, since the exponents have their origin in different physics.   
We remark that this equality was also found to hold in the random
transverse Ising chain \cite{young} and in the random transverse 
Ising spin glass \cite{rieger}.

By calculating the participation ratio we were able to show that the
excitations in the
Bose glass phase are fully localized. This is in contrast to the random
transverse Ising systems, since in these models rare strongly 
coupled {\em clusters} of spins act collectively giving rise to a
divergent susceptibility. Furthermore we find the participation
to obey a scaling relation.

The investigation of the phase transition at the tip of the lobe for
integer boson densities showed
a {\em direct} Mott insulator to superconductor transition for weak 
disorder. The critical exponents we obtain are $z=1, \nu =0.7\pm 0.1$ and
$\eta =0.1 \pm 0.1$ which agree with the exponents of the
pure 3D XY model, so the universality class of the transition
is not altered by weak disorder. This conclusion is supported
by recent mean field renormalization group studies and scaling arguments put
forward in \cite{pazmandi}.

\section{Acknowledgements}
We would like to thank F. Pazmandi, G. Zimanyi and A. P. Young for useful
discussions and the Center of Parallel Computing (ZPR) in
K\"oln for the generous allocation of computing time. This work was
supported by the Deutsche Forschungsgemeinschaft (DFG) and performed within 
the Sonderforschungsbereich (SFB) 341 K\"oln-Aachen-J\"ulich.

\begin{table}

\begin{tabular}{lcl|lcl} \hline
\multicolumn{3}{c}{d. B-H-M} & \multicolumn{3}{c}{RTIC} \\ \hline\hline

Mott insulator & $\chi_{loc}$ & finite & Paramagnetic phase & $\chi_{loc}$
& finite \\ 
 & $\rho$ & = 0 & & m & = 0 \\ \hline
Bose glass & $\chi_{loc}$ & div. & Griffiths phase & $\chi_{loc}$ & div. \\
 & $\rho$ & = 0 &  & m & = 0 \\ \hline
superfluid/ & $\chi_{loc}$ & div. & Ferromagnetic phase & $\chi_{loc}$ &
div. \\
superconducting & $\rho$ & $\neq$ 0 & & m & $\neq$ 0 \\
 & \multicolumn{2}{c|}{"ordered"} & & \multicolumn{2}{c}{"ordered"} \\ \hline

\end{tabular}
\caption{Corresponding phases in the disordered boson Hubbard model (d. B-H-M)
and the random transverse Ising chain (RTIC). $\rho$ is the superfluid
stiffness, $m$ the spontaneous magnetization and $\chi_{loc}$ the
local susceptibility.}
\label{tabdict}
\end{table}

\end{document}